# Exchange Field Induced Magnetoresistance in Colossal Magnetoresistance Manganites

I.N.Krivorotov, K.R.Nikolaev, A.Yu.Dobin, A.M.Goldman, and E.Dan Dahlberg

*School of Physics and Astronomy, University of Minnesota, 116 Church St. SE, Minneapolis, MN 55455*

The effect of an exchange field on electrical transport in thin films of metallic ferromagnetic manganites has been investigated. The exchange field was induced both by direct exchange coupling in a ferromagnet/antiferromagnet multilayer and by indirect exchange interaction in a ferromagnet/paramagnet superlattice. The electrical resistance of the manganite layers was found to be determined by the absolute value of the vector sum of the effective exchange field and the external magnetic field.

Perovskite manganites have recently attracted a lot of attention because their resistance strongly depends on applied magnetic field; an effect known as colossal magnetoresistance (CMR) [1]. In this Letter, we demonstrate that the resistance of thin CMR films depends not only on external magnetic field but also on the effective exchange field that is quantum mechanical in origin. In particular, it is shown that the resistance of a manganite film is determined by the absolute value of the vector sum of the effective exchange field and the external magnetic field.

We have measured the magnetoresistance of two kinds of magnetic multilayer systems involving thin ferromagnetic manganite films. The first system is an antiferromagnetic/ ferromagnetic/ antiferromagnetic (AF/F/AF) trilayer where the F film is a metallic ferromagnet $La_{2/3}Ca_{1/3}(Sr_{1/3})MnO_3$ and the AF films are insulating antiferromagnets $La_{1/3}Ca_{2/3}MnO_3$. The exchange field in this system is created by direct exchange coupling (exchange bias) between the F and the AF layers [2, 3]. The second system is a superlattice consisting of alternating ferromagnetic and paramagnetic metallic layers $(F/P)_N$ where the F layers are $La_{2/3}Ba_{1/3}MnO_3$ manganite films and the P layers are $LaNiO_3$ nickelate films [4]. The F layers are antiferromagnetically coupled via the Ruderman-Kittel-Kasuya-Yosida (RKKY) interaction that creates an indirect exchange field acting on the ferromagnetic manganite layers [5].

All the samples investigated were single-crystal multilayers grown by ozone-assisted molecular beam epitaxy on $SrTiO_3$ (100) substrates. The details of sample preparation and characterization have been given elsewhere [3, 4]. The magnetoresistance of these samples was measured by a four-point ac technique with the current flowing in the plane of the film along the [001] crystallographic direction. A magnetic field of constant magnitude was rotated through 360° in the plane of the sample and the angular dependence of the resistance referred to as anisotropic magnetoresistance (AMR) was measured. Throughout this paper the measured AMR amplitude for a given magnetic field is defined as the difference between the maximum and the minimum resistance divided by the maximum resistance as the field rotates through 360° in the plane of the film.

For reference, the AMR of a single $La_{2/3}Ca_{1/3}MnO_3$ film of the same thickness (29 Å) as the F film in the AF/F/AF

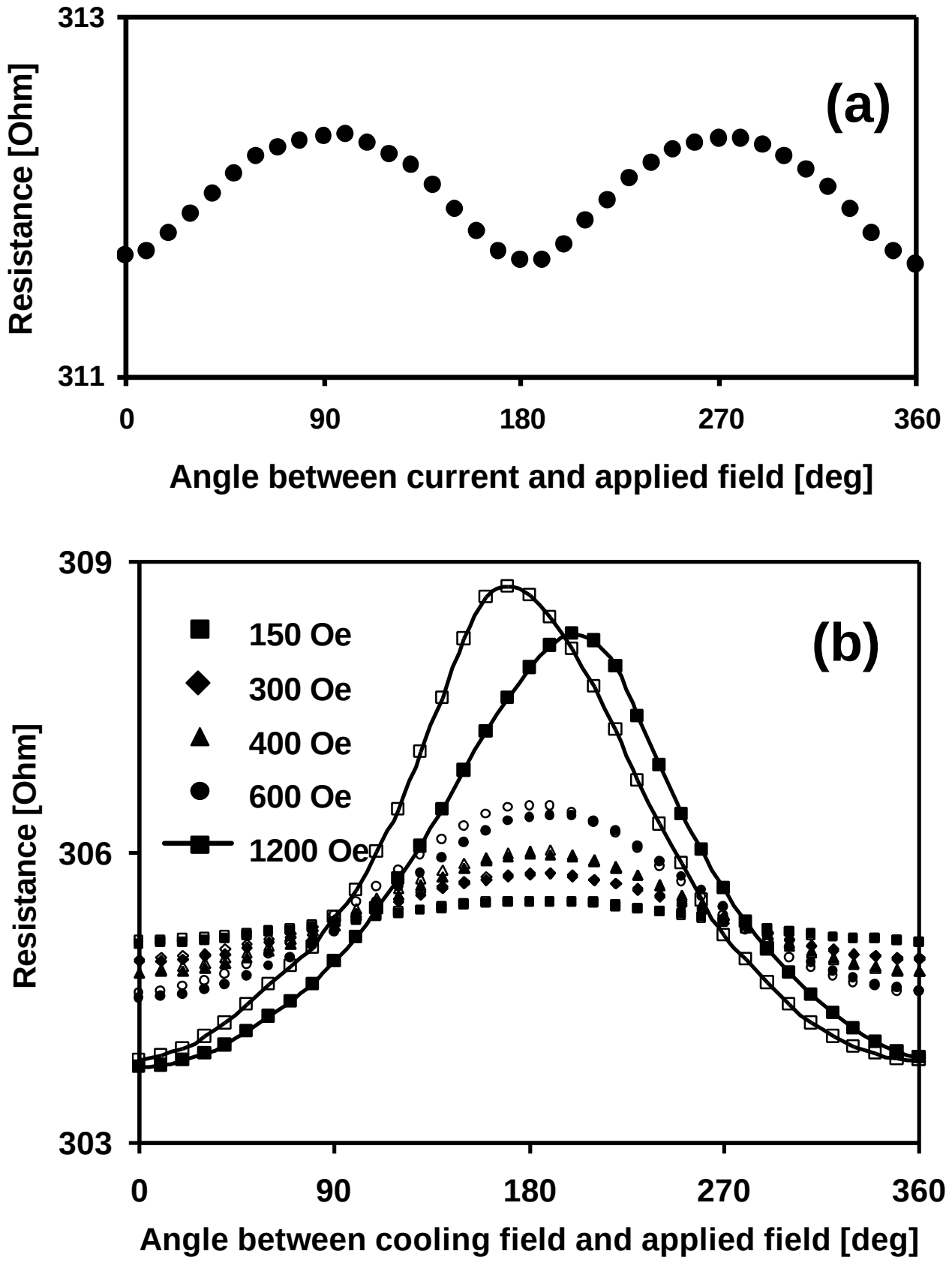

Fig. 1. (a) Resistance versus angle between the current and a saturating applied field of 1 kOe for an unbiased $La_{2/3}Ca_{1/3}MnO_3$ sample at 4.2 K. (b) Dependence of resistance on the angle between the cooling field and the applied field directions for an exchange-biased $La_{2/3}Ca_{1/3}MnO_3$ sample at 4.2 K. Solid symbols represent the data for the clockwise rotation of the applied field while the open symbols represent the data for the counterclockwise rotation.

sample was measured (see Fig.1a). Although AMR in thin films of manganites may significantly differ from the AMR in simple ferromagnets at low temperatures [7], the AMR curves of our ultrathin films have similar shapes to those of transition metal ferromagnets [6]. The reason for this is most probably the small thickness of our films compared to the thickness of those reported in [7]. In particular, (i) the AMR exhibited a 180° periodicity, (ii) sample resistance depended on the angle between the current and magnetization, and (iii) the AMR amplitude first monotonically increased and then saturated with the increasing applied field (see Fig. 2). These three features define the behavior that we will refer to as intrinsic AMR.

The AF/F/AF manganite samples exhibited an effect called exchange bias that results from exchange interaction between magnetic moments of the F and AF layers at their interface [8]. It can be described as arising from an effective exchange field $\mathbf{H}_{\mathbf{EX}}$ acting on the F layer at the AF/F interface or, equivalently, from a unidirectional magnetic anisotropy. The direction of the exchange field is set by the direction of the F magnetization in a saturating field as the AF/F/AF system is cooled to a temperature below the Néel temperature of the AF ($T_N$ = 170 K for $La_{1/3}Ca_{2/3}MnO_3$). The AMR data for a $La_{1/3}Ca_{2/3}MnO_3$(200Å)/ $La_{2/3}Ca_{1/3}MnO_3$(29Å)/ $La_{1/3}Ca_{2/3}MnO_3$(200Å) trilayer are shown in Fig. 1b for several different values of the applied field $H_A$. The AMR periodicity for the exchange biased structure is increased to 360°. The low field curves are fully reproducible for clockwise and counterclockwise rotation of the applied field. However, increasing the field above a threshold leads to the appearance of irreversible changes in $H_{EX}$ [9], which give rise to resistance hysteresis. For large fields ($H_A$ > 2 kOe), the curves exhibit two maxima and the AMR amplitude decreases. Similar results were obtained for an exchange biased $La_{2/3}Sr_{1/3}MnO_3$ sample.

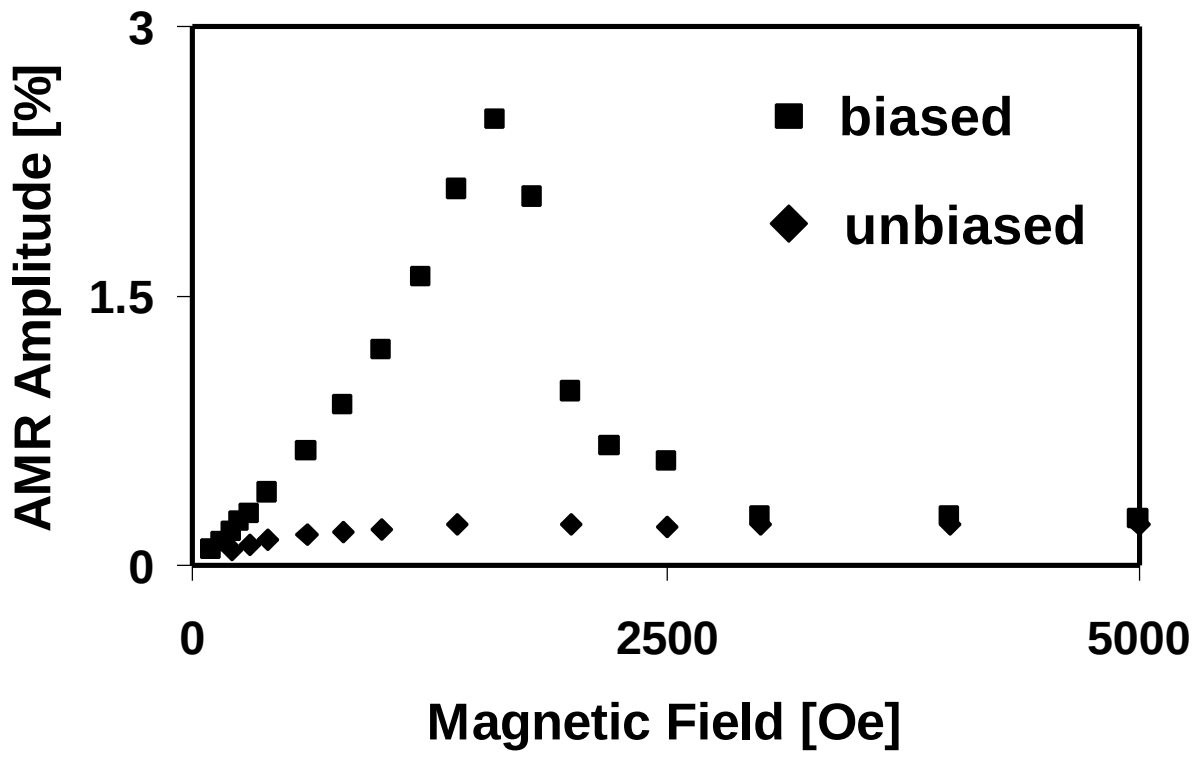

Fig. 2. The AMR amplitude at T = 4.2 K versus applied magnetic field magnitude for $La_{2/3}Ca_{1/3}MnO_3$.

There is a significant qualitative difference between the measured AMR of the exchange biased manganites and conventional exchange biased ferromagnets. The exchange field modifies the measured AMR compared to that of an unbiased film, because magnetization points in direction of the vector sum of the applied field and the exchange field rather then in direction of the applied field. For a conventional ferromagnet, the resistance is determined *only by the direction of its magnetization via the intrinsic AMR effect* and a simple analysis of the angular dependence of the resistance on the in-plane applied field direction yields the exchange field magnitude [10]. The exchange field pins the F layer magnetization thus reducing the amount of the magnetization rotation for modest applied fields. This results in an AMR periodicity of 360° and decreases the AMR amplitude. As the applied field is increased, the degree of rotation increases monotonically, increasing the measured AMR amplitude until the intrinsic AMR amplitude is reached (and the AMR periodicity crosses over from 360° to 180°). In the case of a conventional F, for no value of the applied field can the measured AMR amplitude exceed the amplitude of the intrinsic AMR.

For the exchange biased manganites, however, the observed AMR behavior is different. First, the measured AMR amplitude is a non-monotonic function of the applied field magnitude with a maximum at $H_A$ = 1.6 kOe as shown in Fig. 2. Second, the AMR amplitude at its maximum value exhibits enhancement of approximately a factor of ten relative to the intrinsic AMR amplitude of an unbiased film. Third, the resistance of the exchange biased film depends mainly on the angle between the cooling field and the applied field rather than on the angle between the current and magnetization. These three observations demonstrate that the AMR of exchange biased manganites does not originate from the intrinsic AMR as in the case of conventional ferromagnets.

To explain this behavior, we propose a simple model that is founded on the equivalence of the exchange field and the applied magnetic field similar to that which leads to the Jaccarino-Peter effect in magnetic superconductors [11]. Figure 3 shows a typical dependence of resistance on the magnitude of the applied field for an unbiased film due to the CMR effect. We argue that the AMR of an exchange biased manganite film can be explained by this dependence. Similar to other manganite films [12], the magneto-resistance (R(0 T)-R(5 T))/R(5 T) of our single F layer $La_{2/3}Ca_{1/3}MnO_3$ film has its peak value of 80% near $T = T_C$ while at T = 4.2 K it is 15 %.

In our model, the measured resistance of the AF/F/AF manganite films depends on the magnitude, $H_T$, of the vector sum of the applied magnetic field $\mathbf{H}_{\mathbf{A}}$ and the exchange field $\mathbf{H}_{\mathbf{EX}}$ arising from the two AF layers (see the inset in Fig.3). In the absence of the external applied field (**$H_A$ = 0**), **$H_T$** = **$H_{EX}$** and the resistance of the exchange biased sample is $R_O$. The application of a small external magnetic field **$H_A$** in the opposite direction to **$H_{EX}$** results in decrease of the total field ($H_T = |H_{EX} - H_A|$) and the resistance of the sample increases ($R = R_A$ in Fig.3). Correspondingly, if a small field **$H_A$** is applied in the same direction as **$H_{EX}$** then the total field increases ($H_T = H_{EX} + H_A$) and the resistance decreases ($R = R_B$ in Fig.3).

This model explains the single resistance maximum of the AMR in small fields, as well as the increase of its amplitude with increasing field. The shapes of the AMR curves exhibiting a relatively sharp maximum for applied fields close to the $H_{EX}$ are also well described. The model also explains why the resistance depends on the angle between the applied and exchange field and is almost independent on the current direction in the low field regime. For $H_A > H_{EX}$, the model predicts a decreasing AMR amplitude. Although the simple application of the model for large fields is complicated by the irreversible changes in $H_{EX}$, it is expected that as the applied field increases, the observed AMR will crossover to the intrinsic AMR behavior both in periodicity and magnitude. Therefore, the magneto-transport anisotropy in exchange-biased manganites has its origin in the intrinsic *isotropic* magnetoresistance of the ferromagnet (CMR effect), and can be much larger than the intrinsic AMR of the film. Apparently, it would be observed even if the resistivity of the material did not at all depend intrinsically on the direction of magnetization vector.

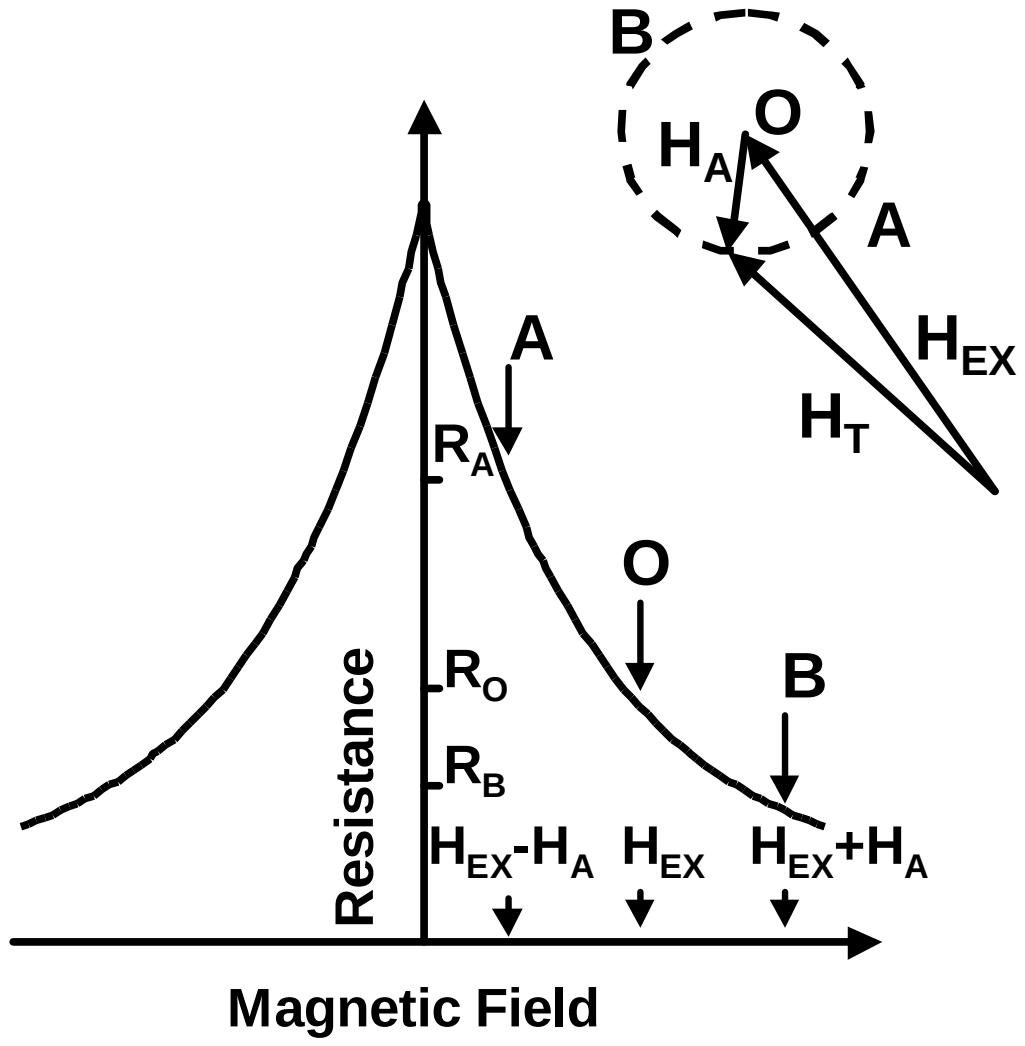

Fig. 3. A typical dependence of resistance on magnetic field for an unbiased manganite film due to the CMR effect. Inset: total field **$H_T$** is a vector sum of the applied magnetic field **$H_A$** and the intrinsic exchange field **$H_{EX}$**.

We have also observed exchange field dependent magnetoresistance in manganite/nickelate superlattices. It has been shown that for thin nickelate spacer layers (3 and 4 unit cells) the manganite layers are coupled antiferromagnetically, while for thicker nickelate spacers the coupling first becomes ferromagnetic and then vanishes [5]. This oscillatory coupling is well described by a simple RKKY theory employing an *ab initio* calculated band structure of nickelate and strong electron scattering in the nickelate spacer layer [5]. The hysteresis loop and magnetoresistance of a $[La_{2/3}Ba_{1/3}MnO_3$ (10 u.c.) / $LaNiO_3$ (4 u.c.)$]_{12}$ superlattice at T = 5 K are shown in Figures 4a and 4b respectively and are found to be almost independent on the in-plane filed direction. The low field magnetoresistance of this AF coupled superlattice is positive in contrast to the negative giant magnetoresistance observed in AF coupled superlattices of transition metal ferromagnets [13]. For higher applied fields, the magnetoresistance becomes negative. In contrast, ferromagnetically coupled and uncoupled superlattices as well as single films of manganites exhibit only negative magnetoresistance such as shown in Fig. 3.

The antiferromagnetic RKKY coupling in the manganite/nickelate superlattices can be represented by effective exchange field **$H_{EX}$** of constant magnitude acting upon each manganite layer in the superlattice. Two manganite layers adjacent to a given layer produce this exchange field that points in the direction opposite to magnetization of these two adjacent layers. We assume, as in the case of the direct exchange field in the AF/F/AF trilayer, that this indirect exchange field alters the resistance of the manganite layer. Again, resistance is a monotonically decreasing function of the absolute value of the total field, which is a vector sum of the applied magnetic field, and the indirect exchange field. In this case, the total field is given by

$$H_T = \sqrt{H_{EX}^2 + H_A^2 - 2H_{EX}H_A m(H_A)} \qquad (1)$$

where $m(H_A)$ is normalized projection of the superlattice magnetization onto the applied field direction ($m(H_A) = M_Z(H_A)/M_S$; $M_Z(H_A)$ is shown in Fig. 4a, $M_S$ is the saturation magnetization of the superlattice). Using Eqn.(1), the $R(H_T)$ shown in Fig. 3 and the experimentally measured $m(H_A)$, we can calculate the magnetoresistance of the AF coupled superlattice with the magnitude of the exchange field $H_{EX}$ as a fitting parameter. The calculated magnetoresistance curve for $H_{EX}$ = 4.3 kOe is shown in Fig. 4c. It is in a good qualitative agreement with the experimentally measured magnetoresistance curve shown in Fig. 4b. We also note that the positions of the resistance maxima occur at the minima of $H_T$ and, therefore, are independent on the exact shape of the magnetoresistance curve shown in Fig. 3 as long as the magnetoresistance is negative. Similar results were obtained for a superlattice with a 3 u.c. thick nickelate spacer layer. The resistance maxima in this superlattice occur at higher values of the applied field that is consistent with a larger AF coupling constant observed for a 3 u.c. thick spacer [5].

This exchange field induced magnetoresistance allows us to extract the approximate magnitude of the exchange field from the magnetoresistance data both in the exchange biased structure and in the AF coupled superlattice. The resistance of the exchange biased film is the largest if the magnitude of the applied field is equal to that of the exchange field with the applied field in the opposite direction to the exchange field leading to the total field being zero. Thus the field at which the largest AMR is observed gives the approximate magnitude of the exchange field. Hysteresis loop measurements revealed a small in-plain biaxial anisotropy for the unbiased film. However, it was found to be much smaller than exchange anisotropy and, therefore, it may be neglected in the analysis of magnetoresistance data In a temperature range from 4.2 K to 170 K, the exchange field determined by the hysteresis loop method (0.6 kOe at T = 4.2 K) was smaller than that given by this magnetoresistance technique (1.6 kOe at T = 4.2 K). This is not surprising, as a similar result was found in conventional exchange biased systems such as Co/CoO bilayers [10]. This is because the reversal of the ferromagnet magnetization significantly decreases the initial magnitude of the exchange field in exchange-biased structures. Therefore, the hysteresis loop shift does not give the *initial* value of the exchange field. On the contrary, AMR measurements employing only a small rotation of the magnetization away from the exchange field direction, do not alter the exchange field and can yield its initial value.

However, in the case of exchange biased manganite films, there is an alternative explanation for the large difference between the hysteresis loop shift and exchange field obtained from the magnetoresistance data. It has been argued that the *low temperature* magnetoresistance in CMR manganite films is a surface or an interfacial effect [12] since it is not observed in bulk manganites. This means that although the whole CMR film is magnetic, only a thin layer near the interface is magnetoresistive. We also note that the direct exchange field, being an interfacial effect, is strongest near the interface and rapidly decays in the interior of the CMR film. The combination of these two factors may be responsible for the larger exchange field obtained from the magnetoresistance data than from the hysteresis loop data. Indeed, magnetoresistance measurements probe only the total field in the thin magnetoresistive layer near the interface where the exchange field is the largest. The exchange field extracted from the hysteresis loop data is an average of the microscopic exchange field taken over the entire sample thickness.

Magnetoresistance of the manganite/nickelate superlattices also supports the interfacial character of the low temperature magnetoresistance in manganite films. The hystereresis loop of the AF coupled superlattice shown in Fig. 4a cannot be described by bilinear antiferromagnetic coupling alone. Biquadratic term also has to be included in the analysis [5]. For this reason, it is difficult to extract the exact value of the exchange field from the hysteresis loop data.

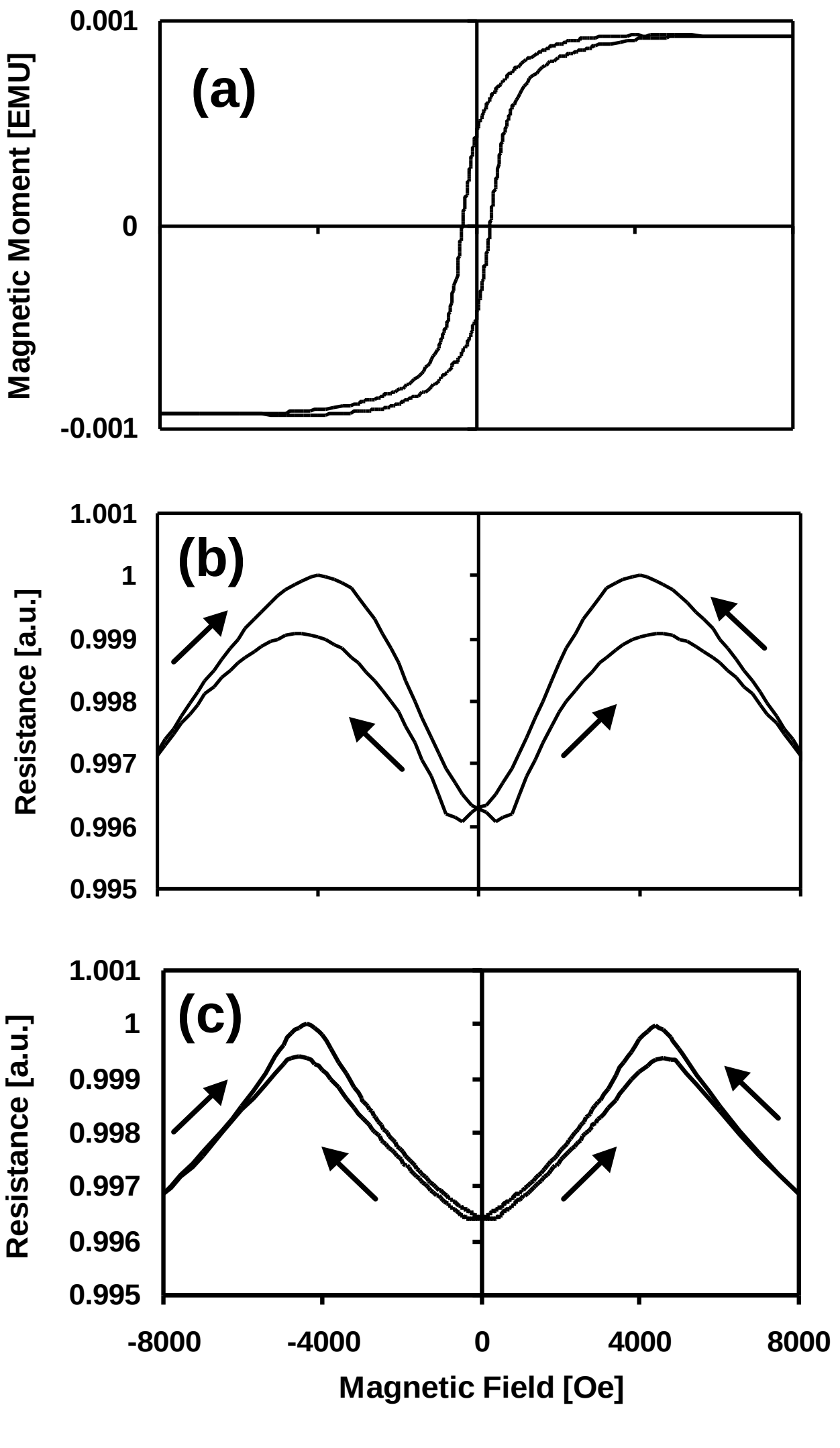

Fig. 4. (a) Hysteresis loop of the antiferromagnetically coupled $[La_{2/3}Ba_{1/3}MnO_3$ (10 u.c.) / $LaNiO_3$ (4 u.c.)$]_{12}$ superlattice at T = 5 K. (b) Resistance versus in-plane magnetic field for this superlattice at T = 5 K. (c) Model prediction for the resistance versus applied field calculated from Eqn. (1) with $H_{EX}$ = 4.3 kOe.

However, the approximate value of the exchange field (proportional to the bilinear coupling constant) obtained from the hysteresis loop is $H_{EX}$ = 1.5 kOe which is significantly less than 4.3 kOe obtained from the magnetoresistance data. In this case, the indirect exchange field is also strongest near the manganite film interface due to the very short mean free path of the conduction electrons in manganites (~ 7 Å as given by the Drude formula). Therefore, our data support the interfacial origin of the low temperature magnetoresistance in thin films of manganites.

In conclusion, the quantum mechanical exchange field alters the resistance of thin films of metallic ferromagnetic manganites. The electrical resistance of these manganite films is determined by the vector sum of the effective exchange field and the external magnetic field. In addition, the magnetoresistance data support the interfacial origin of the low temperature magnetoresistance in single-crystalline thin films of these materials.

This work was supported by the University of Minnesota NSF MRSEC under Grant NSF/DMR-9809364.